\documentclass[aps,prb,reprint,twocolumn,superscriptaddress,amsmath,amssymb,10pt]{revtex4-2}
\usepackage{graphicx}
\usepackage{xcolor}
\usepackage[colorlinks, linkcolor=red,citecolor=blue,urlcolor=blue]{hyperref}
\usepackage[all]{hypcap} 
\begin{document}

\title{Impurity-induced Inverse Faraday Effect}

\author{A.~A. Kopasov}
\affiliation{National University of Science and Technology ``MISIS'', Moscow 119049, Russia}
\author{A.~A.~Bespalov}
\affiliation{Institute for Physics of Microstructures, Russian Academy of Sciences, 603950 Nizhny Novgorod, GSP-105, Russia}
\affiliation{Moscow Institute of Physics and Technology, Dolgoprudnyi, Moscow Region 141701, Russia}
\affiliation{National Research Lobachevsky State University of Nizhny Novgorod, 603950 Nizhny Novgorod, Russia}
\author{A.~S.~Mel'nikov}
\affiliation{Moscow Institute of Physics and Technology, Dolgoprudnyi, Moscow Region 141701, Russia}
\affiliation{Institute for Physics of Microstructures, Russian Academy of Sciences, 603950 Nizhny Novgorod, GSP-105, Russia}
\affiliation{National Research Lobachevsky State University of Nizhny Novgorod, 603950 Nizhny Novgorod, Russia}

\begin{abstract}
We provide a quantum-mechanical description of the photoinduced dc current states and magnetic fields around nonmagnetic point impurities in a two-dimensional (2D) electron gas irradiated by a circularly polarized electromagnetic wave.
Based on the solution of the corresponding time-dependent Schr\"{o}dinger equation within the second-order perturbation theory in the electromagnetic wave amplitude we find that the resulting dc magnetic field component perpendicular to the plane of the 2D system is distributed like in a set of random magnetic fluxes bound to the positions of impurities. As a result, the spatially averaged dc magnetic field does not vanish far from the sample edges and, thus, our scenario of the inverse Faraday effect in disordered systems differs strongly from the standard one based on the relaxation time approximation within the hydrodynamic or kinetic equation approaches which would give only the photoinduced  currents flowing along the sample edges. The predicted mechanism for formation of rectified currents flowing around the impurity centers is shown to be generic both for 2D and three-dimensional systems. The internal dc magnetic field can give rise to the photoinduced Hall effect and Faraday rotation for a probing electromagnetic signal. 
\end{abstract}

\maketitle

There is a growing interest in the field of optoelectronics with the aim of generation and control of the current states and magnetization in solids due to a nonlinear interaction of light with the orbital and spin degrees of freedom of electronic systems~\cite{KimelN2004,KimelN2005,KirilyukRMP2010,Subkhangulov,GhamsariNP2016}. One of the hallmarks of the optoelectronics is the inverse Faraday effect (IFE) --- the generation of the dc magnetic moment $\mathbf{M}^{{\rm dc}}$ in the sample as a result of the transfer of the angular momentum of a circularly polarized electromagnetic wave to the electronic system. This effect was first predicted by Pitaevskii~\cite{Pitaevskii} and then observed in Eu$^{+3}$:CaF garnet~\cite{vanderZiel}. Theoretical description of IFE in metal plasma was developed in Refs.~\cite{HertelJMMM2006,HerterlPRB2015,BattiatoPRB2014}, IFE for a 3D "harmonic atom" model was studied in Ref.~\cite{TokmanPLA1999}, the photoinduced magnetic moment in 2D semiconductor isotropic and anisotropic quantum dots, rings and 1D nanotubes was examined in~\cite{MagarillChaplik}, the magnetic moment produced by twisted light in bulk semiconductors was calculated in Ref.~\cite{QuinteiroEPL2009}, and persistent currents due to the IFE were treated in Refs.~\cite{KoshelevPRB2015,KibisPRL2011,KibisPRB2013,KibisPRB2021}. The well-known expression for the induced \textit{dc magnetic moment} in the collisionless metal plasma reads~\cite{HertelJMMM2006}
\begin{equation}\label{Hertel_result}
 \mathbf{M}^{{\rm dc}} = \frac{ine^3V}{4m_e^2\omega^3 c}\left(\mathbf{E}_0\times\mathbf{E}_0^*\right) \ .
\end{equation}
where $\mathbf{E}_0$ is the complex amplitude of the electric field $\mathbf{E}(t) = \frac{1}{2}\mathbf{E}_0e^{-i\omega t} + {\rm c.c.}$, $\omega$ is the radiation frequency, $e$ is the electron charge, $m_e$ is the effective mass, $c$ is the speed of light, $n$ is the electron concentration, and $V$ is the volume of the sample.

\begin{figure}[htpb]
\centering
\includegraphics[scale = 0.28]{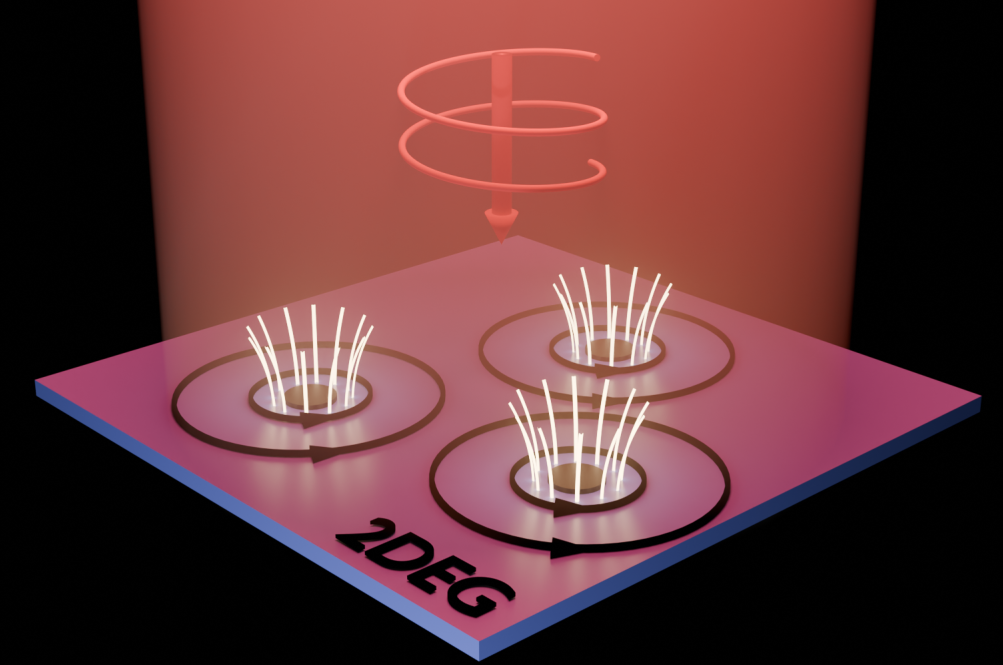}
\caption{Schematic picture illustrating the suggested mechanism of the inverse Faraday effect. We show the lines of the dc currents flowing around each impurity center and magnetic field generated by a circularly polarized light incident on the two-dimensional electron gas (2DEG).}
\label{Fig:Fig1}
\end{figure}

In a standard scenario of the inverse Faraday effect, which is based on the relaxation time approximation within the hydrodynamic~\cite{KarpmanJPP1982,StefanPLA1982} or kinetic equation approaches~\cite{TokmanPRB2020}, the appearance of the dc magnetic field in the sample is explained due to the emergence of the photoinduced dc currents flowing \textit{only} along the sample edges. Direct consequence of such mechanism is that for rather large samples the emergent magnetic field should decrease strongly towards the center of the sample. This fact, however, is in obvious contradiction with recent experimental results~\cite{HanNC2023} demonstrating the presence of huge photoinduced magnetic fields in irradiated graphene discs, which significantly exceed typical magnetic fields generated due to the edge currents. Existing theoretical studies suggest that the photoinduced current and IFE can be dramatically enhanced by plasmonic effects, e.g., by placing a periodic array of nanospheres in the vicinity of the two-dimensional electron liquid~\cite{PotashinPRB2020}. Alternative explanation of IFE enhancement is the generation of a pump-induced synthetic magnetic field~\cite{DurnevPRB2023}. Corresponding calculations performed within the kinetic equation approach reveal a prominent role of the energy dependence of scattering times in the enhancement of the photoinduced Faraday rotation angle.

The above mentioned experimental results motivated us to consider the effects of disorder in the physics of the photoinduced currents and IFE. Note that a similar problem also arises in studies of other optoelectronic effects such as the surface photogalvanic effect and the photon drag~\cite{IvchenkoBook,MagarillJETP1981,AlperovichJETP1981,GurevichPRB1993,MikheevSR2018,KarchPRL2011,CandussioPRB2020,Plank2DM2019,DurnevAP2019,DurnevPSS2020}. 
The simplest generalization of Eq.~(\ref{Hertel_result}) accounting for the disorder effects can be obtained following the Drude-type theory [see, e.g., Refs.~\cite{DurnevPSS2020, GunyagaPRB2023, SharmaPRB2024}]:
\begin{equation}\label{Hertel_disordered_result}
 \mathbf{M}^{{\rm dc}} = \frac{ine^3V\tau^2}{4m_e^2\omega(1+\omega^2\tau^2) c}\left(\mathbf{E}_0\times\mathbf{E}_0^*\right) \ ,
\end{equation}
where $\tau$ is the mean free time between collisions. The magnetic moment is generated by the rectified currents flowing along the edge of the illuminated region (where the electric field is inhomogeneous, so that $\mathop{\mathrm{rot}} \mathbf{M} \neq 0$) or along the sample edge if the whole sample is illuminated. These currents produce a magnetic field that decays towards the center of the illuminated part of the sample. Note also that besides the rectified current states the second-order response should also contain the induced currents at double frequency. These currents are generally present in systems with broken inversion symmetry and can generate inhomogeneous ac magnetization in the sample as it happens in the presence of structured light (see, e.g., Refs.~\cite{GunyagaPRB2023,GunyagaPRL2025}). While experimental studies of the second harmonic generation can provide important insights into the mechanism of the second-order response, in the present work we focus exclusively on the dc part of the response.

The key point of our work is that we suggest an alternative mechanism of generation of dc currents which appear to be nonzero even far from the edges. 
According to this mechanism, the combined effect of the circularly polarized light and the scattering of electrons at each individual impurity generates an additional dc current flowing around the impurity center [see Fig.~\ref{Fig:Fig1}].
This current and related magnetic moment do not appear in the above expression~(\ref{Hertel_disordered_result}) as they manifest the mesoscopic scenario of the transfer of the angular momentum from the incident electromagnetic wave to the electronic system. Such rectified currents, in turn, can generate a nonvanishing magnetic field in the bulk of the sample and far from the sample edges. As far as we know the description of this mechanism is lacking in the literature. We believe that the main reason lies in the fact that the standard theoretical approaches used to tackle the disordered systems usually operate with physical quantities averaged on a scale given by the electron mean free path. Thus, the information about the dc current distribution on a mesoscopic level (on a scale given by a typical distance between the defects) is absent in the averaged theory. The main goal of our work is to fill this gap and carry out the analysis of  the impurity-induced dc current states accounting for the mesoscopic effects described above.

For this purpose we focus on the case of the IFE in the two-dimensional (2D) electron gas and start from the calculation of the photoinduced currents and magnetic moment generated around a single nonmagnetic point impurity. We concentrate on the orbital mechanism for generation of photoinduced currents and disregard the spin-orbit effects associated with broken mirror symmetry of the system~\cite{EdelsteinPRL1998,TanakaNJP2020}. Based on the solution of the time-dependent Schr\"{o}dinger equation within the second-order perturbation theory over the electric field, we calculate the time-averaged angular momentum transferred to the electronic system by a circularly polarized electromagnetic wave. We demonstrate that the induced dc magnetic moment component perpendicular to the plane of the 2D system ($z$-projection) $M_z$ exhibits oscillatory behavior at a length scale $v_F/\omega$ ($v_F$ is the Fermi velocity) with the average
growing linearly with increasing distance from the defect center $r$. The corresponding dc current density also oscillates and its average decays as 
\begin{equation}\label{current_density_main_result}
j_{\varphi}(r) = Q/r^2 
\end{equation}
far away from the defect center ($r\gg v_F/\omega$). The factor $Q$ is determined by the characteristics of the incident electromagnetic wave, 2D electron gas, and the impurity potential
\begin{equation}\label{Q_main_result}
Q = \left(e^3 E_0^2/4\pi^2m_e^2\omega^3\right)k_F^2\sigma(\epsilon_F) \ .
\end{equation}
Here $\sigma(\epsilon_F)$ is the scattering cross section at the Fermi energy $\epsilon_F = \hbar^2k_F^2/2m_e$. It is instructive to note that the resulting profile of the current density~(\ref{current_density_main_result}) is similar to the spatial profile of the supercurrent around a Pearl vortex in a thin superconducting film far from the vortex center~\cite{deGennesBook,AbrikosovBook}. Thus, the solution of the Maxwell equations with the current distribution~(\ref{current_density_main_result}) yields that the averaged magnetic field component perpendicular to the plane of the 2D system $B_z^{\rm dc}$ is distributed like in a magnetic flux tube (of the radius $\sim v_F/\omega$) centered at the position of the defect center $B_z^{\rm dc} = \phi\delta(\mathbf{r})$, where $\phi = 4\pi^2Q/c$ is the magnetic flux generated by a single impurity. In a sample with a random impurity array the above mentioned features of the current distribution should lead to a random distribution of the induced dc magnetic field [see Fig.~\ref{Fig:Fig1}]. We find that the resulting oscillations of the magnetic field associated with random impurity positions are averaged out in the high-frequency limit and the remaining average field  takes the form $\bar{B}_z \sim \phi n_{\rm imp}$. Here $n_{\rm imp}$ is the impurity concentration. We find that the above mechanism for formation of recitified currents flowing around the impurity centers is generic both for 2D and three-dimensional (3D) case.

We support the above qualitative picture by performing quantum mechanical calculations. For definiteness, we consider the degenerate non-interacting 2D electron gas with quadratic energy-momentum relation. A circularly polarized electromagnetic wave propagates along the $z$-axis, and its characteristic length scale in the plane of the 2D gas is assumed to be much larger than the size of the sample. Note that the presence of nanoscale field inhomogeneities can modify the dc current states due to the change in the electric field distribution acting on the conduction electrons. Provided the electromagnetic radiation illuminates a certain finite area, the field gradients at the boundary of this area will cause the generation of additional dc photocurrents and the size of the illuminated area will play the role of the effective sample size. These currents will circulate in the opposite direction comparing to the currents around the impurities inside the illuminated region partially weakening, thus, the impurity-induced dc magnetic field. In this work we analyze the steady-state response of the electronic system to the driving field, so, the electric field is adiabatically switched on from a distant past (from $t \to -\infty$). The electric field is described by homogeneous vector potential
\begin{equation}\label{vector_potential}
 \mathbf{A}(t) = (cE_0/\omega)\left[\cos(\omega t)\mathbf{e}_x + \sin(\omega t)\mathbf{e}_y\right]\exp(\delta t) \ ,
\end{equation}
where $E_0$ is the field amplitude, $\mathbf{e}_{x}$, $\mathbf{e}_y$, and $\mathbf{e}_z$ are the unit vectors in Cartesian coordinates, $\omega$ is the radiation frequency, and $\delta$ is an infinitesimally small positive number. Dynamics of the electronic states is governed by the time-dependent Schr\"{o}dinger equation
\begin{equation}\label{time_dependent_SE_full}
 i\hbar\frac{\partial\Psi}{\partial t} = \left[\frac{{(\hat{\mathbf{p}}}-(e/c)\mathbf{A})^2}{2m_e} + U(\mathbf{r})\right]\Psi(\mathbf{r},t) \ .
\end{equation}
Here $\hat{\mathbf{p}} = -i\hbar\nabla$ is the canonical momentum operator. For simplicity, we assume that the range $a$ of the impurity potential $U(\mathbf{r})$ is much smaller than $k_F^{-1}$, so that it can be considered as a delta-potential $U(\mathbf{r}) = U_0a^2\delta(\mathbf{r})$. It is important to note that the impurity potential is treated exactly. In 2D systems the delta-potential can be characterized by the following asymptotic behavior of the wave function at $r\to 0$~\cite{SolvableModels}:
\begin{equation}\label{asymptotic_condition}
\Psi(\mathbf{r},t) \approx \Psi_0(t)\ln(R_0/r) \ ,
\end{equation}
where $\Psi_0(t)$ is an arbitrary function and $R_0$ is a positive constant (see discussion in Supplementary Material~\cite{SM}). Here $\mathbf{r}=(r,\varphi)$ is the polar coordinate system with the origin at the impurity position. The point defect has an energy-dependent scattering cross-section
\begin{equation}
\sigma(\epsilon_k) = 4k^{-1}/[1 + (4/\pi^2)\ln^2(kr_0)] \ ,
\end{equation}
where the wave number $k$ and energy are related via $\epsilon_k = \hbar^2k^2/2m_e$, $r_0 = R_0e^{\gamma}/2$, and $\gamma \approx 0.577$ is the Euler-Mascheroni constant. The scattering phase shift $\alpha$ is, in turn, given by the relation $\cot(\alpha) = (2/\pi)\ln(kr_0)$. Note that within the Born approximation $\sigma(\epsilon_k)= (1/4)k^3a^4(U_0/\epsilon_k)^2$.

Our goal is to calculate the time-averaged part of the $z$-projection of the induced magnetization
\begin{equation}\label{magnetic_moment_definition}
 \mathcal{M}_z(r,t) = (e/2m_ec) L_z(r, t)  \ .
\end{equation}
Here 
\begin{eqnarray}\label{angular_momentum_definition}
 \mathbf{L}(r, t) = \ \  \\
 \nonumber
= 2\int \frac{d\varphi}{2\pi}\sum_{\lambda} f_{\lambda} \Psi_{\lambda}^*(\mathbf{r},t)(\mathbf{r}\times [\hat{\mathbf{p}}-(e/c)\mathbf{A}])\Psi_{\lambda}(\mathbf{r},t) 
\end{eqnarray}
is the averaged over $\varphi$ angular momentum density transferred to the electronic system by an electromagnetic wave. Here $\lambda$ denotes the set of quantum numbers, $f_{\lambda}$ is the distribution function in a distant past (of the Fermi-Dirac form), and the factor of two in Eq.~(\ref{angular_momentum_definition}) accounts for the spin degeneracy.

\begin{figure*}[htpb]
\centering
\includegraphics[scale = 0.75]{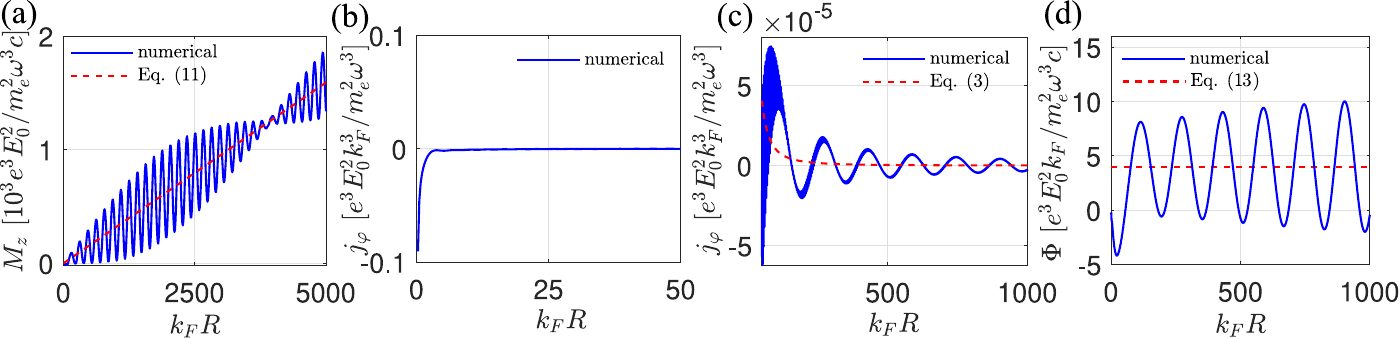}
\caption{(a) Typical $M_z(R)$ dependence. (b)-(c) Typical plots of the dc current density $j_{\varphi}(R)$, $k_FR$ varies within the range $k_FR\in [0, 50]$ for (b) and $k_FR\in[50, 1000]$ for (c). (d) Magnetic flux $\Phi$ through a circular area of radius $R$ centered at the impurity. Here solid lines show the results of numerical integration of analytical expressions~\cite{SM}, whereas the dashed lines in (a), (c), and (d) are the results of Eq.~(\ref{2D_magnetic_moment_analytical}), (\ref{current_density_main_result}), and (\ref{averaged_flux}), respectively. We choose $\hbar\omega = 0.08\epsilon_F$ and $k_Fr_0 = 1$ to produce the plots.}
\label{Fig:results_2D}
\end{figure*}

Details of the solution of Eq. \eqref{time_dependent_SE_full} can be found in the Supplementary Material~\cite{SM}. First, using the gauge transformation, we eliminate the term propotional to $\mathbf{A}^2(t)$ from Eq.~\eqref{time_dependent_SE_full}, and then we obtain a solution of the resulting equation up to quadratic terms in $E_0$. Using the second-order perturbation theory we disregard, of course, the electronic spectrum modifications arising from the higher order nonlinearities  in the electric field amplitude~\cite{KibisPRB2020,BoevSR2023,KibisJPCM2024}. Next, we calculate the magnetic moment $M_z(R)$ defined as the magnetization integrated over the circle of the radius $R$ around the impurity (see~\cite{SM} for corresponding cumbersome expressions), the dc current density $j_{\varphi}(R)$ as well as the magnetic flux $\Phi(R)$ through a circular area of the radius $R$ centered at the impurity position. Typical results for zero temperature are shown in Fig.~\ref{Fig:results_2D}. Note also that the described mechanism for the generation of dc magnetization is applicable only for conducting systems. IFE in insulating systems appears due to other mechanisms \cite{NovaNatPhys2016,BanerjeePRB2022,GribovaJETPL2024}.

One can see that the resulting dc magnetic moment contains the fast oscillating contribution at the length scale $ v_F/\omega$ arising from the step-like behavior of the distribution function. The envelope of these oscillations, in turn, varies at the length scale $k_F(v_F/\omega)^2$. Note that both these length scales have been previously observed in quantum mechanical calculations of the edge photogalvanic effect~\cite{BespalovRQE2025}. Averaging out these oscillations and taking the limit $\hbar\omega \ll E_F$ and $R\gg v_F/\omega$, we find the following expression for the dc magnetic moment: 
\begin{equation}\label{2D_magnetic_moment_analytical}
 M_z =  \left(e^3E_0^2/m_e^2\omega^3c\right)\left(k_F^2R/4\pi\right)\sigma(\epsilon_F) \ .
\end{equation}
The linear asymptotic behavior of the magnetic moment~(\ref{2D_magnetic_moment_analytical}) as a function of $R$  
is in a good agreement with the results of numerical integration [see Fig.~\ref{Fig:results_2D}]. It is important to note that the sign of the magnetic moment~(\ref{2D_magnetic_moment_analytical}) appears to be opposite to the one generated by the sample edge currents [see Eq.~(\ref{Hertel_result})]. On a qualitative level this sign difference can be explained if we consider a sample with a defect in the form of a hole and apply the Hertel-type theory~\cite{HertelJMMM2006}. 
In this theory, the current is given by the curl of the magnetization, and its direction is set by the cross product of the magnetization and the outward normal to the boundary. Since the outward normals at the sample edge and at the hole perimeter point in opposite directions, the resulting rectified currents will have opposite signs.

Let us also brieflly discuss the relevance of the obtained results to the case of more realistic models of the impurity potential. We verified that Eqs.~(\ref{current_density_main_result}), (\ref{Q_main_result}), and (\ref{2D_magnetic_moment_analytical}) are also valid for the screened Coulomb potential within the first Born approximation, for which $\sigma(\epsilon_F) \sim k_F^{-1}$ (see~\cite{SM} for details). In contrast, our assumption of the impurity as a predominantly isotropic scatterer is obviously violated in the case of the unscreened Coulomb potential, which requires a separate treatment.

The magnetic moment described by Eq.~(\ref{2D_magnetic_moment_analytical}) is generated by the current density given by the expression~(\ref{current_density_main_result}).  
To find the corresponding magnetic field we solve the Maxwell equation (details of the derivation are provided in~\cite{SM}) 
\begin{equation}
{\rm curl} \ {\rm curl}\mathbf{A} = (4\pi/c)\mathbf{j}(\mathbf{r})\delta(z) \ .
\end{equation}
and obtain $\mathbf{B}(\mathbf{r},z = 0) = \phi \mathbf{e}_z \delta(\mathbf{r})$, where 
\begin{equation}\label{averaged_flux}
\phi = \left(e^3E_0^2/m_e^2\omega^3c\right)k_F^2\sigma(\epsilon_F) \ .
\end{equation}
Keeping in mind that the solution~(\ref{2D_magnetic_moment_analytical}) has been obtained in the limit $r\gg v_F/\omega$, we conclude that the width of the magnetic field peak should be of the order of $v_F/\omega$.

Taking now a random 2D impurity array, one can see that the problem of calculation of dc photoinduced magnetic fields can be mapped to the one describing an array of Pearl vortices in a superconducting film (in the limit of vanishing effective penetration depth~\cite{deGennesBook,AbrikosovBook}). 
The currents induced by a given impurity appear to be compensated by the ones generated by other scatterers at the scale of the inter-defect distance. The magnetic moments of individual impurities fluctuate due to a random distance between scattering centers. 
The resulting distribution of the dc current and magnetic field in a sample should depend on the statistics of the inter-defect distances.
Provided both the average inter-defect distance and its root mean square deviation $\delta r$ are larger than the oscillation length scale $v_F/\omega$ the fluctuations of the magnetic field are averaged out (see Fig.~\ref{Fig:results_2D}(d)) and the expression for the spatially averaged magnetic field can be cast to the form
\begin{equation}
\bar{B}_z = \left(e^3E_0^2/m_e^2\omega^3c\right)k_F^2\sigma(\epsilon_F)n_{\rm imp} \ .
\end{equation}
We can compare the field $\bar{B}_z$ with the field $B_z^{\rm dc}$ associated with edge currents. For a sample (e.g., a disk) with characteristic size $R_{\rm S}$ we have $B_z^{\rm dc} \sim M_z^{\rm dc}/R_{\rm S}^3$ in the center of the sample, so that $B_z^{\rm dc}/\bar{B}_z \sim l/R_{\rm S}$. Thus, the impurity-induced field should dominate for samples that are much larger than the mean free path. The relative strength of the predicted phenomena is larger in samples with smaller mean-free paths. 
Taking, for instance, the parameters corresponding to the bilayer graphene (see, e.g.~Refs~\cite{CandussioPRB2020, GunyagaPRB2023} and references therein) - the electron concentration $n = 10^{12}$~cm$^{-2}$, the effective mass $m_e \sim 0.01m_0$, where $m_0$ is the free electron mass, the radiation frequency $\omega/2\pi = 1$~THz, the electromagnetic wave intensity $\sim 10^2$~$\mu$W/$\mu$m$^2$, and $k_Fl\gtrsim 1$, we get the averaged magnetic field $\bar{B}_z \sim 10^2$~G. This value can be even larger for frequency ranges in the vicinity of a plasmonic resonance. 
Although we restricted ourselves to the case of zero temperature, we verified that the resulting average dc current and magnetic flux are not affected by temperature in the limit $T\ll \epsilon_F$, whereas the amplitudes of the oscillations on $M_z(R)$, $j_{\varphi}(R)$, and $\Phi(R)$ curves decrease with the temperature increase and become smeared out in the case $T\gg \hbar\omega$. 

It is important to note that that our results are not restricted to 2D systems. Calculations of the induced dc magnetic moment, rectified current density, and the induced dc magnetic field in 3D case for a single impurity are provided in~\cite{SM}. Below we summarize our findings. We choose the coordinate system with the impurity at the origin and the axes directions consistent with Eq.~(\ref{vector_potential}).
We obtain the following expression for the averaged dc magnetization integrated over a sphere with radius $R$ around the defect center:
\begin{equation}
 M_z = \frac{e^3E_0^2}{m_e^2\omega^3c}\frac{k_F^3R\sigma_{\rm 3D}(\epsilon_F)}{6\pi^2} \ , 
\end{equation}
where $\sigma_{\rm 3D}(\epsilon_F)$ is the scattering cross-section at the Fermi level in 3D case and $E_0$ is the amplitude of the field acting on the electrons near the defect center. Corresponding expression for the averaged dc current density has the form
\begin{equation}
\mathbf{j}(\mathbf{r}) = Q_{\rm 3D}\frac{\mathbf{e}_z\times \mathbf{r}}{r^4} \ , \ \ \ \ Q_{\rm 3D} = \frac{e^3E_0^2k_F^3\sigma(\epsilon_F)}{8\pi^3m_e^2\omega^3} \ .
\end{equation}
Here $\mathbf{r} = x\mathbf{e}_x + y\mathbf{e}_y + z\mathbf{e}_z$. This rectified current generates the magnetic field distribution $\mathbf{B} = (4\pi Q_{\rm 3D}/c)z\mathbf{r}/r^4$ and the magnetic flux though the plane perpendicular to the $z$-axis $\phi = 4\pi^2Q_{\rm 3D}/c$.

Let us also note that the electron-electron interaction can modify the predicted dc current states through several mechanisms. First, static screening of the impurity potential can alter the scattering cross-section, affecting, thus, the prefactor $Q$ in our expressions. Second, the radiation field around each impurity can induce variations in in the electric charge distribution, which should be determined self-consistently. However, this effect is expected to be weak provided the spatial scales of the current states (e.g., $v_F/\omega$) significantly exceed the screening length. Third, the radiation field can also renormalize the defect potential, leading to higher-order nonlinear corrections in the electric field amplitude that are beyond the scope of this study. Finally, inelastic electron–electron scattering and related broadening effects may smear the oscillatory terms in our results without substantially modifying the slowly decaying current profiles.

To sum up, we suggest a generic impurity-induced mechanism of the inverse Faraday effect. Considering the case of nonmagnetic point impurities, we have shown that the circular polarized electromagnetic radiation induces dc current-carrying states around each impurity position. The corresponding photoinduced magnetic fields are distributed like in a set of random magnetic fluxes and the resulting averaged magnetic field does not vanish even far from the sample edges. We expect that these magnetic fields can provide important contributions to the photoinduced Hall and Kerr effects.

\acknowledgments

The authors thank O.~Kibis, V.~Kovalev, M.~Durnev and V.~Kachorovskii for stimulating and useful discussions.

This work was supported  by the Federal Academic Leadership Program Priority 2030 (NUST MISIS Strategic Technology Project "Quantum Internet") in part of quantum-mechanical calculations of the magnetic moment, by the Russian Science Foundation (Grant No. 25-12-00042) in part of the analysis of electrodynamics of the considered systems, and by the Grant of the Ministry of science and higher education of the Russian Federation No. 075-15-2025-010 in part of numerical calculations. A.A.K. acknowledges the financial support of the Foundation for the Advancement of Theoretical Physics and Mathematics BASIS (Grant No. 23-1-2-32).

The data that support the findings of this article are openly available~\cite{Zenodo}.

\clearpage
\onecolumngrid

\setcounter{section}{0}
\setcounter{equation}{0}
\setcounter{figure}{0}
\setcounter{table}{0}
\renewcommand{\theequation}{\Alph{section}\arabic{equation}}
\renewcommand{\thefigure}{S\arabic{figure}}
\renewcommand{\thetable}{S\arabic{table}}

\begin{center}
{\large\bfseries Supplementary Material: Impurity-induced Inverse Faraday Effect\par}
\vspace{0.8em}
{A.~A. Kopasov,\textsuperscript{1}
A.~A. Bespalov,\textsuperscript{2,3,4} and
A.~S. Mel'nikov\textsuperscript{3,2,4}\par}
\vspace{0.6em}
{\small
\textsuperscript{1}National University of Science and Technology ``MISIS'', Moscow 119049, Russia\\
\textsuperscript{2}Institute for Physics of Microstructures, Russian Academy of Sciences, 603950 Nizhny Novgorod, GSP-105, Russia\\
\textsuperscript{3}Moscow Institute of Physics and Technology, Dolgoprudnyi, Moscow Region 141701, Russia\\
\textsuperscript{4}National Research Lobachevsky State University of Nizhny Novgorod, 603950 Nizhny Novgorod, Russia
}
\end{center}
\vspace{1.0em}
\appendix

Here we provide the detailed derivation of our main results. Our starting point is the Schr\"{o}dinger equation~(6) in the main text with the vector potential $\mathbf{A}(t) = (cE_0/\omega)\left[\cos(\omega t)\mathbf{e}_x + \sin(\omega t)\mathbf{e}_y\right]e^{\delta t}$, which is solved within the second-order perturbation theory over the electric field. The obtained solutions are then used for the calculations of the dc quantities, such as the $z$-projection of magnetization, the current density component $j_{\varphi}$, and the magnetic flux generated by a single impurity. 

Supplementary material is organized as follows. In Sec.~\ref{boundary_conditions_section} we discuss the boundary conditions  for a point impurity. In Sec.~\ref{screened_potential} we obtain an estimate for the scattering cross section of electrons by a screened Coulomb potential in a 2D electron liquid. In Sec.~\ref{wave_function_corrections} we obtain a perturbative solution of the Schr\"odinger equation. In Sec.~\ref{magnetization} we present the expressions for $M_z(R)$, $j_{\varphi}(R)$, and $\Phi(R)$ used to produce the plots in Fig.~2 in the main text and also provide the derivation of the induced dc magnetic field for averaged current density given by Eq.~(3) in the main text. Sections~\ref{boundary_conditions_section}, \ref{screened_potential}, \ref{wave_function_corrections}, and~\ref{magnetization} correspond to the two-dimensional case. Finally, we present our results on the induced dc magnetization, rectified current density, and the induced dc magnetic field for the three-dimensional case in Sec.~\ref{3D_case}.

\section{Boundary conditions for a point impurity}\label{boundary_conditions_section}

In this section, we discuss the origin of the boundary condition (7) and the connection between $R_0$ and characteristics of the impurity potential.

Let us obtain a solution of the Schr\"{o}dinger equation (6) close to the impurity, assuming that the characteristic electron energy $\epsilon$ satisfies the relation
\begin{equation}
	|\epsilon| \ll \frac{\hbar^2}{2m_ea^2},
	\label{eq:E_small}
\end{equation}
where $a$ is the range of the potential $U(\mathbf{r})$. If we keep only the leading terms in Eq. (6) for $r \lesssim a$, we obtain
\begin{equation}
  \left[-\frac{\hbar^2\nabla^2}{2m_e} + U(\mathbf{r})\right]\Psi(\mathbf{r},t) = 0.
	\label{eq:short_Schrodinger}
\end{equation}
For $r>a$, the solution of Eq. \eqref{eq:short_Schrodinger} has the form (7) with some contributions with non-zero angular momentum projection, which vanish in the limit $r \to 0$. $\Psi_0(t)$ can take any value due to the linearity of Eq. \eqref{eq:short_Schrodinger}, and $R_0$ is determined by $U(\mathbf{r})$. As an example, let us consider a rectangular potential
\begin{equation}
	U(\mathbf{r}) = \begin{cases}
		U_0 & \mbox{when } r<a, \\
		0 & \mbox{when } r\geq a.
	\end{cases}
	\label{eq:U_rectangular}
\end{equation}
We will find the solution of Eq. \eqref{eq:short_Schrodinger} with zero angular momentum projection (we assume that solutions with non-zero projections are weakly affected by the short-range potentials). If $U_0 < 0$, for $r<a$ we have
\begin{equation}
	\Psi(\mathbf{r},t) = \Psi_1(t) J_0(\kappa r),
	\label{eq:Psi_inside}
\end{equation}
where $J_n$ is the Bessel function of the first kind, $\Psi_1(t)$ is an arbitrary function, and $\kappa = (2m_e |U_0|/\hbar^2)^{1/2}$. Using the continuity of $\Psi$ and $\partial \Psi/\partial r$ at $r=a$, we obtain
\begin{equation}
	R_0 = a \exp \left( \frac{J_0(\kappa a)}{\kappa a J_1(\kappa a)} \right).
	\label{eq:R0_well}
\end{equation}
Clearly, $R_0$ can take any non-negative value due to the oscillating nature of the Bessel functions. 

It should be noted that the boundary condition (7) is responsible for an impurity-induced bound state with energy
\begin{equation}
	\epsilon_b  = -\frac{\hbar^2}{2m_e r_0^2}
	\label{eq:epsilon_b}
\end{equation}
and wave function
\begin{equation}
	\Psi_b = \mathrm{const} \cdot K_0(\kappa_b r),
	\label{eq:kappa_b}
\end{equation}
where $\kappa_b = (2m_e |\epsilon_b|/\hbar^2)^{1/2}$, and $K_m$ denotes a modified Bessel function of the second kind~\cite{Abramowitz}. The existence of bound states is natural when $U(\mathbf{r})$ corresponds to a potential well.

The analog of Eq. \eqref{eq:R0_well} for $U_0 > 0$ is
\begin{equation}
	R_0 = a \exp \left( -\frac{I_0(\kappa a)}{\kappa a I_1(\kappa a)} \right),
	\label{eq:R0_barrier}
\end{equation}
where $I_m$ are modified Bessel function of the first kind~\cite{Abramowitz}. One can see that $R_0 < a$ in this case. Though we strictly derived this constraint on $R_0$ for a model potential, we claim that the condition $R_0 \lesssim a$ should be valid for a potential barrier ($U(\mathbf{r}) >0$) in the general case. Indeed, if the opposite condition $R_0 \gg a$ would be fullfilled we would find that the applicability condition \eqref{eq:E_small} of the point potential approximation is satisfied for $\epsilon = \epsilon_b$, and we arrive at the unphysical conclusion that a potential barrier induces a bound state.

\section{Scattering cross section for a screened Coulomb potential} \label{screened_potential}
In this section we will analyze the scattering cross section for the screened potential of a point charge in a 2D electron liquid. First, we calculate the screened potential $\varphi_e$ using Lindhard's approach \cite{Electron_liquid}. The potential satisfies the equation
\begin{equation}
	-\nabla^2 \varphi_e = 4\pi [\rho_{\mathrm{ext}}(\mathbf{r}) + \delta(z) \rho_{\mathrm{ind}}(x,y)],
	\label{eq:Poisson}
\end{equation}
where $\rho_{\mathrm{ext}}(\mathbf{r})$ is the external charge density, and $\rho_{\mathrm{ind}}$ is the induced charge density due to the redistribution of electrons in the plane of the electron liquid (charge per unit area). We consider a point external charge: $\rho_{\mathrm{ext}}(\mathbf{r}) = q_{\mathrm{ext}} \delta(\mathbf{r})$. Applying the two-dimensional Fourier transform to Eq. \eqref{eq:Poisson} yields
\begin{equation}
	k^2 \varphi_e(\mathbf{k},z) - \frac{\partial^2}{\partial z^2} \varphi_e(\mathbf{k},z) = 4\pi [q_{\mathrm{ext}} + \rho_{\mathrm{ind}}(\mathbf{k})] \delta(z),
	\label{eq:Poisson_Fourier}
\end{equation}
where
\begin{equation}
	\varphi_e(\mathbf{k},z) = \int \!\!\! \int \varphi_e(\mathbf{r}) \exp(-ik_x x -ik_y y) dx \, dy, \quad \rho_{\mathrm{ind}}(\mathbf{k}) = \int \!\!\! \int \rho_{\mathrm{ind}} (x,y) \exp(-ik_x x -ik_y y) dx \, dy.
	\label{eq:phi&rho_Fourier}
\end{equation}
The induced charge density is related to the in-plane potential through the density-density susceptibility $\chi_{nn}(\mathbf{k})$:
\begin{equation}
	\rho_{\mathrm{ind}}(\mathbf{k}) = e^2 \chi_{nn}(\mathbf{k}) \varphi_e(\mathbf{k},0).
	\label{eq:rho_phi}
\end{equation}
For a 2D electron liquid at zero temperature the susceptibility is \cite[Table 4.2]{Electron_liquid}
\begin{equation}
	\chi_{nn}(\mathbf{k}) = 
	\begin{cases}
		- \nu & \mbox{for } k<2k_F, \\
		-\nu \left( 1 - \frac{\sqrt{k^2 - 4k_F^2}}{k} \right) & \mbox{for } k>2k_F,
	\end{cases}
	\label{eq:chi_nn}
\end{equation}
where $\nu = m/(\pi \hbar^2)$ is the electronic density of states in two dimensions. Equations \eqref{eq:Poisson_Fourier} and \eqref{eq:rho_phi} yield
\begin{equation}
	\varphi_e(\mathbf{k},z) = \frac{2\pi q_{\mathrm{ext}}}{k - 2\pi e^2 \chi_{nn}(\mathbf{k})} e^{-k|z|}.
	\label{eq:phi_solution}
\end{equation}
The Fourier transform of the potential energy of an electron is
\begin{equation}
	\tilde{U}(k) = e\varphi_e(\mathbf{k},0) = 
	\begin{cases}
		\frac{2\pi eq_{\mathrm{ext}}}{k + \lambda^{-1}_{\mathrm{TF}}} & \mbox{for } k<2k_F, \\
		\frac{2\pi eq_{\mathrm{ext}}}{k + \lambda^{-1}_{\mathrm{TF}}\left( 1 - \frac{\sqrt{k^2 - 4k_F^2}}{k} \right)} & \mbox{for } k>2k_F,
	\end{cases}
	\label{eq:screened_potetial}
\end{equation}
where $\lambda_{\mathrm{TF}} = (2\pi e^2 \nu)^{-1}$ is the Thomas-Fermi screening length. We can calculate the scattering cross section for this potential profile within the first Born approximation. The scattering amplitude for electrons with energy $\epsilon_F$ is proportional to $\tilde{U}(2k_F \sin(\theta/2))$, where $\theta$ is the scattering angle. It is clear that when the Fermi wavelength $\lambda_F = 2\pi/k_F$ is much larger than $\lambda_{\mathrm{TF}}$, the scattering is almost isotropic. The scattering cross section is then
\begin{equation}
	\sigma(\epsilon_F) \approx \frac{m^2}{\hbar^4 k_F} |\tilde{U}(0)|^2 = \frac{\pi q_{\mathrm{ext}}^2}{2e^2} \lambda_F.
	\label{eq:cross_section}
\end{equation}
For $q_\mathrm{ext} = e$ we obtain $\sigma(\epsilon_F) = \pi \lambda_F/2$, which exceeds the maximum possible $s$-wave scattering cross section $2\lambda_F/\pi$. This indicates that the parameters used lie outside the applicability domain of the first Born approximation, which actually works for $|q_\mathrm{ext}| \ll |e|$. We believe that our calculations at least yield a valid order-of-magnitude estimate of the scattering cross section for a charged impurity: $\sigma(\epsilon_F) \sim \lambda_F$.

\section{Perturbative solution of the Schr\"odinger equation}
\label{wave_function_corrections}

In this section we outline how the perturbative solution of Eq. (6) with the boundary condition (7) can be obtained. We also provide a list of corrections to the unperturbed wave functions which are relevant for the calculations of the magnetic moment and magnetic field.

First, we eliminate the term $\propto \mathbf{A}^2(t)$ form Eq.~(6), which is responsible for a time-dependent global phase of the wave function and is irrelevant for the calculations of the angular momentum~(10) and magnetization~(9). Substitution
\begin{equation}
 \Psi(\mathbf{r},t) \to \Psi(\mathbf{r},t)\exp\left[-\frac{i}{\hbar}\int_{-\infty}^td\tau \frac{e^2}{2m_ec^2}\mathbf{A}^2(\tau)\right]
\end{equation}
into Eq.~(6) yields a simplified equation
\begin{equation}\label{time_dependent_SE_reduced}
 i\hbar\frac{\partial\Psi}{\partial t} = \left[-\frac{\hbar^2\nabla^2}{2m_e} + \frac{i\hbar e}{m_ec}\mathbf{A}(t)\nabla + U(\mathbf{r})\right]\Psi(\mathbf{r},t) \ .
\end{equation}
This equation is solved within the second-order perturbation theory over the electric field for all relevant initial states $\Psi_{m,k}^{(0)}(\mathbf{r},t)$, which are cylindrical waves characterized by the wave number $k$ and $z$-projection of the angular momentum $\hbar m$ ($m$ is an integer). In the present work we restrict ourselves to the contribution of the continuum states to the angular momentum and neglect possible contribution of the localized bound states, which appear for an attractive potential $U$.

The initial states (existing before the field is switched on) with nonzero $z$-projections of the angular momentum are unaffected by the condition~(7):
\begin{equation}\label{2D_nonzero_angular_momentum_normalized}
 \Psi_{m \neq 0,k}^{(0)}(\mathbf{r},t) = \sqrt{k/2\pi}J_{|m|}(kr)e^{im\varphi}e^{-i\epsilon_k t/\hbar} \ ,
\end{equation}
whereas the states with zero projection take the form
\begin{equation}\label{2D_zero_angular_momentum_normalized}
 \Psi^{(0)}_{m=0,k}(\mathbf{r},t) = \frac{\left[\frac{2}{\pi}\ln(kr_0)J_0(kr) - Y_0(kr)\right]}{\sqrt{\frac{2\pi}{k}\left[1 + \frac{4}{\pi^2}\ln^2(kr_0)\right]}}e^{-i\epsilon_k t/\hbar} \ .
\end{equation}
Here $Y_m(x)$ is the Bessel function of the second kind, and we employ the normalization condition
\begin{equation}
 \int d^2\mathbf{r}\Psi_{m,k}^{(0)*}(\mathbf{r},t)\Psi_{m',k'}^{(0)}(\mathbf{r},t) = \delta_{m,m'}\delta(k-k') \ .
\end{equation}

As a next step, we find the relevant perturbation corrections to the initial states~(\ref{2D_nonzero_angular_momentum_normalized}) and (\ref{2D_zero_angular_momentum_normalized}) up to the second-order terms in the electric field. The time-averaged current density vanishes in the absence of the impurity. This means that all contributions to the current density that do not contain $r_0$ will cancel each other out, and hence we do not need to take them into account. As a result, within the second order in $E_0$ we need to calculate only the wave functions with projections of the angular momentum $m=0,\pm 1$. In the following we denote the first- (second-) order correction for the initial state $\Psi_{m,k}^{(0)}$ as $\Psi_{m,k}^{(1)}$ $\left(\Psi_{m,k}^{(2)}\right)$. Note that, of course, the time and spatial dependencies of the corrections are different in comparison with the corresponding initial state. The relevant first-order corrections over $E_0$ can be presented in the form (we omit the normalization factor $\sqrt{k/(2\pi)}$ for brevity)
\begin{subequations}\label{first_order_corrections}
 \begin{align}
  \Psi_{0,k}^{(1)}(\mathbf{r},t) = \frac{-ieE_0k}{2m_e\omega^2}e^{-i\epsilon_k t/\hbar}e^{i\omega t-i\varphi}C_0(k)\left[-\frac{2}{\pi}\ln(kr_0)J_1(kr) + Y_1(kr) + q^{(1)}_{0,k}\right]\\
  \nonumber
  + \frac{ieE_0k}{2m_e\omega^2}e^{-i\epsilon_kt/\hbar}e^{-i\omega t + i\varphi}C_0(k)
  \times\left[-\frac{2}{\pi}\ln(kr_0)J_1(kr)+Y_1(kr) + \frac{ik_+}{k}H_1^{(1)}(k_+r)\right] \ ,\\
 \Psi_{1,k}^{(1)}(\mathbf{r},t) = 
	\frac{ieE_0 k}{2m_e \omega} e^{\delta t - i\epsilon_k t/\hbar} \left[ - \frac{J_0(kr) + q_{1,k}^{(1)}(r)}{\omega - i\delta} e^{i\omega t} + \frac{J_2 (kr)}{\omega + i\delta} e^{2i\varphi -i\omega t} \right] \ ,\\
 \Psi_{-1,k}^{(1)}(\mathbf{r},t) 
	= \frac{ieE_0 k}{2m_e\omega} e^{\delta t -i\epsilon_k t/\hbar} \left\{ \frac{1}{\omega+i \delta} \left[ J_0(kr) - \frac{H_0^{(1)}(k_+ r)}{1 + i \frac{2}{\pi} \ln(k_+ r_0)} \right] e^{-i\omega t} + \frac{J_2(kr)}{\omega -i\delta} e^{i\omega t - 2i\varphi} \right\} \ .
 \end{align}
\end{subequations}
We have to keep $\delta$ in the expressions for $\Psi_{1,k}^{(1)}$ and $\Psi_{-1,k}^{(1)}$ to obtain the correct second-order corrections, which have the form
\begin{subequations}\label{second_order_corrections}
 \begin{align}
  \Psi_{1,k}^{(2)}(\mathbf{r},t) 
	= \frac{e^2 E_0^2 k}{4m_e^2 \omega^4} \left[ -k J_1(kr) + q_{1,k}^{(2)}(r) \right] e^{-i \epsilon_k t/\hbar +i\varphi} \ , \\
 \Psi_{-1,k}^{(2)}(\mathbf{r},t) 
	= \frac{e^2 E_0^2 k}{4m_e^2 \omega^4} \left[ -k J_1(kr) + \frac{k_+ H_1^{(1)}(k_+ r) - k H_1^{(1)}(kr)}{ 1 + i\frac{2}{\pi} \ln(k_+ r_0)} \right] e^{-i \epsilon_k t/\hbar -i\varphi} \ .
 \end{align}
\end{subequations}
In the above expressions $H^{(1)}_m(x)$ is the Hankel function of the first kind~\cite{Abramowitz}, $k_{\pm} = \sqrt{k^2 \pm 2m_e\omega/\hbar}$, $C_0(k) = 1/\sqrt{1+(4/\pi^2)\ln^2(kr_0)}$, and we also introduced several new functions:
\begin{subequations}\label{2D_various_functions}
 \begin{align}
  q_{0,k}^{(1)} = \begin{cases}\cfrac{ik_-}{k}H_1^{(1)}(k_-r) \ , \ \ \text{for} \ \epsilon_k > \hbar\omega \ ,\\
                   \cfrac{2}{\pi}\cfrac{|k_-|}{k}K_1(|k_-|r) \ , \ \ \text{for} \ \epsilon_k < \hbar\omega \ ;
                  \end{cases} \ \ \ \
  q_{1,k}^{(1)}(r) = \begin{cases}
            \cfrac{-H_0^{(1)}(k_-r)}{1 + i\cfrac{2}{\pi}\ln(k_-r_0)} \ , \ \ \text{for} \ \epsilon_k > \hbar\omega \ ,\\
            \cfrac{K_0(|k_-|r)}{\ln(|k_-|r_0)} \ , \ \ \text{for} \ \epsilon_k < \hbar\omega \ ;
           \end{cases} \\
  q_{1,k}^{(2)}(r) = \begin{cases}\cfrac{k_-H_1^{(1)}(k_-r) - kH_1^{(1)}(kr)}{1 + i\cfrac{2}{\pi}\ln(k_-r_0)} \ , \ \ \text{for} \ \epsilon_k>\hbar\omega \ , \\
        \cfrac{-|k_-|K_1(|k_-|r) + i\cfrac{\pi k}{2}H_1^{(1)}(kr)}{\ln(|k_-|r_0)} \ , \ \ \text{for} \ \epsilon_k < \hbar\omega \ .
                     \end{cases}
 \end{align}
\end{subequations}
Note that we ignored terms with angular momentum projections $m=-1$ and $3$ in the expression for $\Psi_{1,k}^{(2)}$ and terms with $m=-3$ and $m=1$ in the expression for $\Psi_{-1,k}^{(2)}$, because these terms do not contribute to the time-averaged current density within second order in $E_0$. We did not calculate $\Psi_{0,k}^{(2)}$ due to the same reason.

\section{Expressions for the dc magnetization, current density and the magnetic flux. Derivation of the dc magnetic field generated by a single impurity}\label{magnetization}

Here we provide the expressions for $M_z(R)$, $j_{\varphi}(R)$, and $\Phi(R)$ used to produce the plots in Fig.~(2) in the main text and then provide the details of the derivation for Eqs.~(13), (14). Substituting the solutions given by Eqs.~\eqref{2D_nonzero_angular_momentum_normalized}, \eqref{2D_zero_angular_momentum_normalized}, (\ref{first_order_corrections}) and (\ref{second_order_corrections}) into Eq.~(9) for $M_z(t)$ and taking the dc component, we get rather cumbersome expression for the zero-temperature dc magnetization integrated over the circle of radius $R$ around the impurity:
\begin{subequations}\label{2D_magnetic_moment}
 \begin{align}
  M_z(R) = \sum_{m = -1}^1 \left(P_m + Q_m\right) \ ,\\
  P_{0} = -\frac{e^3E_0^2}{2m_e^2\omega^3c}\int_0^{k_F}dk \int_0^R dr \frac{k^2r^2}{1 + \frac{4}{\pi^2}\ln^2(kr_0)}\left[\frac{2}{\pi}\ln(kr_0)J_0(kr)-Y_0(kr)\right]\biggl\{2\left[-\frac{2}{\pi}\ln(kr_0)J_1(kr)+Y_1(kr)\right] \\
  \nonumber
  +{\rm Re}(q_{0,k}^{(1)})-\frac{k_+}{k}Y_1(k_+r)\biggl\} \ ,\\
  P_{1} = -\frac{e^3E_0^2}{2m_e^2\omega^3c}{\rm Re}\int_0^{k_F}dk\int_0^R dr \ k^2r^2J_1(kr)\left[J_0(kr) + q_{1,k}^{(1)}\right] \ ,\\
  P_{-1} = -\frac{e^3E_0^2}{2m_e^2\omega^3c}{\rm Re}\int_0^{k_F}dk\int_0^R dr \ k^2r^2J_1(kr)\left[J_0(kr) -\frac{H_0^{(1)}(k_+r)}{1 + i\frac{2}{\pi}\ln(k_+r)}\right] \ ,\\
  Q_0 = \frac{\hbar e^3E_0^2}{4m_e^3\omega^4c}\int_0^{k_F}dk\int_0^R dr \frac{k^3r}{1 + \frac{4}{\pi^2}\ln^2(kr_0)}\biggl\{\frac{k_+^2}{k^2}|H_1^{(1)}(k_+r)|^2-|q_{0,k}^{(1)}|^2 \\
  \nonumber
  +2\left[\frac{2}{\pi}\ln(kr_0)J_1(kr) - Y_1(kr)\right]\left[\frac{k_+}{k}Y_1(k_+r)+{\rm Re}(q_{0,k}^{(1)})\right]\biggl\} \ ,\\
  Q_1 = \frac{\hbar e^3E_0^2}{2m_e^3\omega^4c}{\rm Re}\int_0^{k_F}dk\int_0^R dr \ k^2r J_1(kr)q_{1,k}^{(2)} \ ,\\
  Q_{-1} = -\frac{\hbar e^3E_0^2}{2m_e^3\omega^4c}{\rm Re}\int_0^{k_F}dk\int_0^R dr \  k^2r J_1(kr)\frac{k_+H_1^{(1)}(k_+r)-kH_1^{(1)}(kr)}{1 + i\frac{2}{\pi}\ln(k_+r_0)} \ .
 \end{align}
\end{subequations}
In order to produce the plots shown in Fig.~2 in the main text, the integrals in the above expressions have been calculated numerically. For the derivation of Eq.~(11) in the main text we replace the integrands in Eqs.~(\ref{2D_magnetic_moment}) with their asymptotic expressions in the large $r$ limit, neglect the oscillating and exponentially decaying contributions, and in the end take the limit $\hbar\omega/E_F\ll1$.  The plots of the dc current density $j_{\varphi}(R)$ shown in Fig.~2(b) and 2(c) in the main text are obtained using the relation $j_{\varphi}(R) = (c/\pi R^2)[\partial M_z(R)/\partial R]$. This expression simply follows from the definition of $M_z(R)$, which is the $z$-projection of the magnetization integrated over a circular area with the radius $R$ around the impurity $M_z(R) = (\pi/c)\int_0^Rj_{\varphi}(r)r^2dr$.  For the calculations of the magnetic flux through a circular area with the radius $R$ around the impurity $\Phi(R)$ it is convenient to substitute the expression for the vector potential in the real-space representation
\begin{equation}
\mathbf{A}(\mathbf{r},z = 0) = \frac{1}{c}\int\frac{\mathbf{j}(\mathbf{r}')}{|\mathbf{r}-\mathbf{r}'|}d^2\mathbf{r}' 
\end{equation}
into the expression for the magnetic flux $\Phi(R) = R\int_0^{2\pi}A_{\varphi}(\mathbf{R})d\varphi$, which can be further cast to the form
\begin{equation}
\Phi(R) = \frac{2\pi}{c}\int\frac{\mathbf{R}\mathbf{r}'}{r'|\mathbf{R}-\mathbf{r}'|}j_{\varphi}(r')d^2\mathbf{r}' \ .
\end{equation}
Subsituting numerically obtained current distribution $j_{\varphi}(r')$ into the above expression and performing the integration over $d^2\mathbf{r}'$ numerically, we obtain $\Phi(R)$. Note that some caution is needed for numerical evaluation of the above integral. To eliminate the singularities in the integrand, it is convenient to add and subtract the auxillary function 
\begin{equation}
\tilde{\Phi}(R) = \frac{2\pi R}{c}j_{\varphi}(R)\int d^2\mathbf{r}' \ \frac{e^{-|\mathbf{R}-\mathbf{r}'|/\varkappa}}{|\mathbf{R}-\mathbf{r}'|} = \frac{4\pi^2R}{c}j_{\varphi}(R) \varkappa \ ,
\end{equation}
where $\varkappa > 0$ is an arbitrary constant. So, the expression for the induced magnetic flux can be rewritten in the following form
\begin{equation}
\Phi(R) = \frac{2\pi}{c}\int d^2\mathbf{r}' \ \left[\frac{\mathbf{R}\mathbf{r}'}{r'}j_{\varphi}(r')-Re^{-|\mathbf{R}-\mathbf{r}'|/\varkappa}j_{\varphi}(R)\right]\frac{1}{|\mathbf{R}-\mathbf{r}'|} + \frac{4\pi^2 R}{c}j_{\varphi}(R)\varkappa \ .
\end{equation}

To find the corresponding magnetic field of the averaged current given by Eq.~(3) in the main text, we solve the Maxwell equation 
\begin{equation}
{\rm curl} \ {\rm curl}\mathbf{A} = (4\pi/c)\mathbf{j}(\mathbf{r})\delta(z) \ .
\end{equation}
Introducing the Fourier transforms of the field and current 
\begin{subequations}
\begin{align}
\mathbf{A}(\mathbf{r}, z) = \int\frac{d^2\mathbf{q}}{(2\pi)^2}\int\frac{dk}{(2\pi)}\mathbf{A}_{\mathbf{q}k}e^{-i(\mathbf{q}\mathbf{r}+kz)} \ ,\\
\mathbf{A}(\mathbf{r}, z = 0) = \int\frac{d^2\mathbf{q}}{(2\pi)^2}\mathbf{A}_{\mathbf{q}}e^{-i\mathbf{q}\mathbf{r}} \ ,\\
\label{j_q_definition}
\mathbf{j}(\mathbf{r}) = \int\frac{d^2\mathbf{q}}{(2\pi)^2}\mathbf{j}_{\mathbf{q}}e^{-i\mathbf{q}\mathbf{r}}  \ ,
\end{align}
\end{subequations}
and taking the Fourier transform of the current density 
\begin{equation}
\mathbf{j}_{\mathbf{q}} = \frac{2\pi Q i(\mathbf{e}_z\times\mathbf{q})_z}{q} \ ,
\end{equation}
one finds the solution in the form~\cite{deGennesBook,AbrikosovBook} 
\begin{subequations}
\begin{align}
\mathbf{A}_{\mathbf{q}k} = \frac{4\pi}{c}\frac{\mathbf{j}_q}{(\mathbf{q}^2 + k^2)} \ , 
(\mathbf{B}_{\mathbf{q}k})_z = -\frac{4\pi}{c}\frac{i(\mathbf{q}\times\mathbf{j}_q)_z}{(q^2 + k^2)} \ ,\\
\mathbf{B}_{\mathbf{q}} = \frac{4\pi^2Q}{c}\mathbf{e}_z \ .
\end{align}
\end{subequations}
In the coordinate space the magnetic field is peaked at the position of impurity:  
$\mathbf{B}(\mathbf{r},z = 0) = \phi \mathbf{e}_z \delta(\mathbf{r})$, where 
\begin{equation}\label{averaged_flux_SM}
\phi = \left(e^3E_0^2/m_e^2\omega^3c\right)k_F^2\sigma(\epsilon_F) \ .
\end{equation}

\section{Induced dc magnetization, rectified current density, and the induced dc magnetic field in the three-dimensional case}\label{3D_case}
Here we present the calculations of the induced dc magnetization in the three-dimensional (3D) case. The impurity is located at $\mathbf{r} = 0$ and we also assume that the $s$-wave scattering cross-section is much larger than scattering cross-sections with higher angular momenta. Hereafter, we use the spherical coordinates $(r,\theta,\varphi)$ and choose the basis of spherical waves for our calculations. Eigenstates of the Hamiltonian for zero external electric field and in the absence of the impurity potential are as follows:
\begin{equation}\label{3D_nonzero_angular_momentum}
 \Psi_{l,m,k}(\mathbf{r},t) = \sqrt{\frac{2}{\pi}}kj_l\left(kr\right)Y_{l,m}(\hat{\mathbf{r}})e^{-i\epsilon_kt/\hbar} \ ,
\end{equation}
where $j_l$ is the spherical Bessel function~\cite{Abramowitz}, $Y_{l,m}(\hat{\mathbf{r}})$ denotes spherical harmonics, $\hat{\mathbf{r}} = \mathbf{r}/r$, $l$ is the orbital momentum quantum number, $m$ is the magnetic quantum number, and $\epsilon_k = \hbar^2k^2/2m_e$. In 3D case with a point impurity the wave function should satisfy the following asymptotic condition at $r \to 0$ \cite{SolvableModels}:
\begin{equation}
	\Psi(\mathbf{r},t) \approx \Psi_0(t) \left( 1 + \frac{r_0}{r} \right),
	\label{boundary_condition_3D}
\end{equation}
where $\Psi_0(t)$ is an arbitrary function, and $r_0$ is a real constant. Correspondingly, for $E_0 = 0$ the eigenstates with $l > 0$ are given by Eq.~(\ref{3D_nonzero_angular_momentum}), while the zero angular-momentum state can be written in the form
\begin{equation}\label{3D_zero_angular_momentum_normalized}
 \Psi_{0,0,k}(\mathbf{r},t) = \sqrt{\frac{2}{\pi}}k\frac{\left[j_0(kr) - kr_0y_0(kr)\right]}{\sqrt{1 + k^2r_0^2}}Y_{0,0}(\hat{\mathbf{r}})e^{-i\epsilon_kt/\hbar} \ .
\end{equation}
For $r_0<0$, there is also a bound state with wave function
\begin{equation}
	\Psi_b^{(3D)} = \frac{1}{\sqrt{2\pi |r_0|}} \frac{e^{r/r_0}}{r} e^{-i \epsilon'_b t/\hbar}
	\label{eq:bound_3D}
\end{equation}
and energy $\epsilon_b' = -\hbar^2/(2mr_0^2)$.
One can verify that all the states including $\Psi_{0,0,k}(\mathbf{r},t)$ and $\Psi_b^{(3D)}(\mathbf{r},t)$ defined above form a complete basis. In further considerations we focus on the case $r_0 > 0$. Note that the point impurity in 3D case has the energy-dependent scattering cross-section of the form
\begin{equation}
\sigma_{\rm 3D}(\epsilon_k) = \frac{4\pi r_0^2}{1 + k^2r_0^2} \ .
\end{equation}

As a next step, we calculate the perturbative corrections for the eigenstates within the second order in $E_0$ using the boundary condition~(\ref{boundary_condition_3D}). Note that the perturbation operator
\begin{eqnarray}
V(\mathbf{r},t) = \frac{i\hbar eE_0}{2m_e\omega} e^{\delta t} \biggl[e^{i(\omega t - \varphi)}\left(\sin\theta\frac{\partial}{\partial r} - \frac{i}{r\sin\theta}\frac{\partial}{\partial\varphi} + \frac{\cos\theta}{r}\frac{\partial}{\partial\theta}\right) \\
\nonumber
+ e^{-i(\omega t - \varphi)}\left(\sin\theta\frac{\partial}{\partial r} + \frac{i}{r\sin\theta}\frac{\partial}{\partial\varphi} + \frac{\cos\theta}{r}\frac{\partial}{\partial\theta}\right)\biggl] \ ,
\end{eqnarray}
acting on the states with the angular structure $Y_{l,m}(\hat{\mathbf{r}})$ induces the following corrections:
\begin{equation}
 V(\mathbf{r},t)Y_{l,m}(\hat{\mathbf{r}}) \rightarrow Y_{l\pm 1, m\pm 1}(\hat{\mathbf{r}}) \ .
\end{equation}
We found that for calculations of the $z$-projection of the time-averaged magnetic moment $M_z^{\rm dc}(R)$ defined as the magnetization integrated over a sphere with a radius $R$ around the impurity, it is sufficient to take only the first-order corrections in $E_0$ for the state with $l = 0$. For the states $l = 1$, $m = \pm 1$ we need the second-order corrections as well (only time-independent contributions are relevant). Irrelevant eigenstates and perturbative corrections not presented below produce contributions to magnetization, which then vanish upon time averaging and/or spatial integration.

Below we provide a list with relevant first-order corrections (we omit the normalization factor $k\sqrt{2/\pi}$ for brevity)
\begin{subequations}
 \begin{align}
  \Psi^{(1)}_{0,0,k}(\mathbf{r},t) = \frac{eE_0}{m_e\omega^2}\frac{1}{\sqrt{6}}\frac{e^{-i\epsilon_k t/\hbar - i\omega t}}{\sqrt{1 + k^2r_0^2}}Y_{1,1}(\hat{\mathbf{r}})k\left[ij_1(kr) - ikr_0y_1(kr) + \frac{k_+^2}{k}r_0h_1^{(1)}(k_+r)\right] + \\
  \nonumber
  +\frac{eE_0}{m_e\omega^2}\frac{1}{\sqrt{6}}\frac{e^{-i\epsilon_k t/\hbar + i\omega t}}{\sqrt{1 + k^2r_0^2}}Y_{1,-1}(\hat{\mathbf{r}})k\left[ij_1(kr)-ikr_0y_1(kr) + q_{0,0,k}(r)\right] \ ,\\
 \Psi^{(1)}_{1,1,k}(\mathbf{r},t) = \frac{eE_0}{m_e\omega}\frac{e^{\delta t -i\epsilon_kt/\hbar - i\omega t}}{\sqrt{5}(\omega+i\delta)}Y_{2,2}(\hat{\mathbf{r}})kij_2(kr) + \frac{eE_0}{m_e\omega}\frac{e^{\delta t -i\epsilon_k t/\hbar + i\omega t}}{\sqrt{30}(\omega-i\delta)}Y_{2,0}(\hat{\mathbf{r}})kij_2(kr) + \\
 \nonumber
 +\frac{eE_0}{m_e\omega}\frac{e^{\delta t -i\epsilon_k t/\hbar + i\omega t}}{\sqrt{6}(\omega-i\delta)}Y_{0,0}(\hat{\mathbf{r}})k\left[ij_0(kr) + q_{1,1,k}(r)\right] \ ,\\
 \Psi_{1,-1,k}^{(1)}(\mathbf{r},t) = \frac{eE_0}{m_e\omega}\frac{e^{\delta t -i\epsilon_kt/\hbar + i\omega t}}{\sqrt{5}(\omega-i\delta)} Y_{2,-2}(\hat{\mathbf{r}})ikj_2(kr) + \frac{eE_0}{m_e\omega}\frac{e^{\delta t -i\epsilon_k t/\hbar - i\omega t}}{\sqrt{30}(\omega+i\delta)}Y_{2,0}(\hat{\mathbf{r}})ikj_2(kr) + \\
 \nonumber
 + \frac{eE_0}{m_e\omega}\frac{e^{\delta t -i\epsilon_k t/\hbar - i\omega t}}{\sqrt{6}(\omega+i\delta)}Y_{0,0}(\hat{\mathbf{r}})k\left[ij_0(kr)-\frac{k_+r_0}{1-ik_+r_0}h_0^{(1)}(k_+r)\right] \ ,\\
 \end{align}
\end{subequations}
Here we introduced
\begin{subequations}
\begin{align}
q_{0,0,k}(r) = \begin{cases} \frac{k_-^2}{k}r_0h_1^{(1)}(k_-r) \ , \ \ {\rm for} \ \epsilon_k > \hbar\omega \ ,\\
                  -i\frac{|k_-|^2}{k}r_0\kappa_1(|k_-|r) \ , \ \ {\rm for} \ \epsilon_k < \hbar\omega \ ,
                 \end{cases} \\
q_{1,1,k}(r) = \begin{cases}\frac{-k_-r_0}{1-ik_-r_0}h_0^{(1)}(k_-r) \ , \ \ \rm{for} \ \epsilon_k >\hbar\omega \ ,\\
                 \frac{i|k_-|r_0}{1 + |k_-|r_0}\kappa_0(|k_-|r) \ , \ \ \rm{for} \ \epsilon_k < \hbar\omega \ ,
                \end{cases} 
\end{align}
\end{subequations}
$y_l$ is the spherical Bessel function of the second kind~\cite{Abramowitz}, $h_l^{(1)} = j_l + iy_l$, and $\kappa_l$ is the modified spherical Bessel function of the third kind~\cite{Abramowitz}. It is important to keep $\delta$ in the expressions for $\Psi^{(1)}_{1,1,k}(\mathbf{r},t)$ and $\Psi_{1,-1,k}^{(1)}(\mathbf{r},t)$ to obtain the correct second-order corrections.

The relevant second-order corrections have the form (we omit the normalization factor $k\sqrt{2/\pi}$ for brevity)
\begin{subequations}
\begin{align}
 \Psi_{1,1,k}^{(2)}(\mathbf{r},t) = \frac{-e^2E_0^2}{m_e^2\omega^4}Y_{1,1}(\hat{\mathbf{r}})e^{-i\epsilon_kt/\hbar} \left[ \frac{k^2}{5} j_1(kr) + \frac{i}{6} q_{1,1,k}^{(2)}(r) \right]\ ,\\
 \Psi_{1,-1,k}^{(2)}(\mathbf{r},t) = \frac{-e^2E_0^2}{m_e^2\omega^4}Y_{1,-1}(\hat{\mathbf{r}})e^{-i\epsilon_k t/\hbar} \left\{ \frac{k^2}{5} j_1(kr) +  \frac{ikr_0}{6\left(1 - ik_+r_0\right)}\left[k_+^2h_1^{(1)}(k_+r)-k^2h_1^{(1)}(kr)\right] \right\} \ ,
\end{align}
\end{subequations}
with 
\begin{equation}
 q_{1,1,k}^{(2)}(r) = \begin{cases}\frac{kr_0}{1-ik_-r_0}\left[k_-^2h_1^{(1)}(k_-r)-k^2h_1^{(1)}(kr)\right] \ , \ \ \rm{for} \ \epsilon_k>\hbar\omega \ ,\\
 \frac{-ikr_0}{1 + |k_-|r_0}\left[|k_-|^2\kappa_1(|k_-|r)-ik^2h_1^{(1)}(kr)\right] \ , \ \ \rm{for} \ \epsilon_k < \hbar\omega \ .
 \end{cases} 
\end{equation}

\begin{figure}[htpb]
\centering
\includegraphics[scale = 0.6]{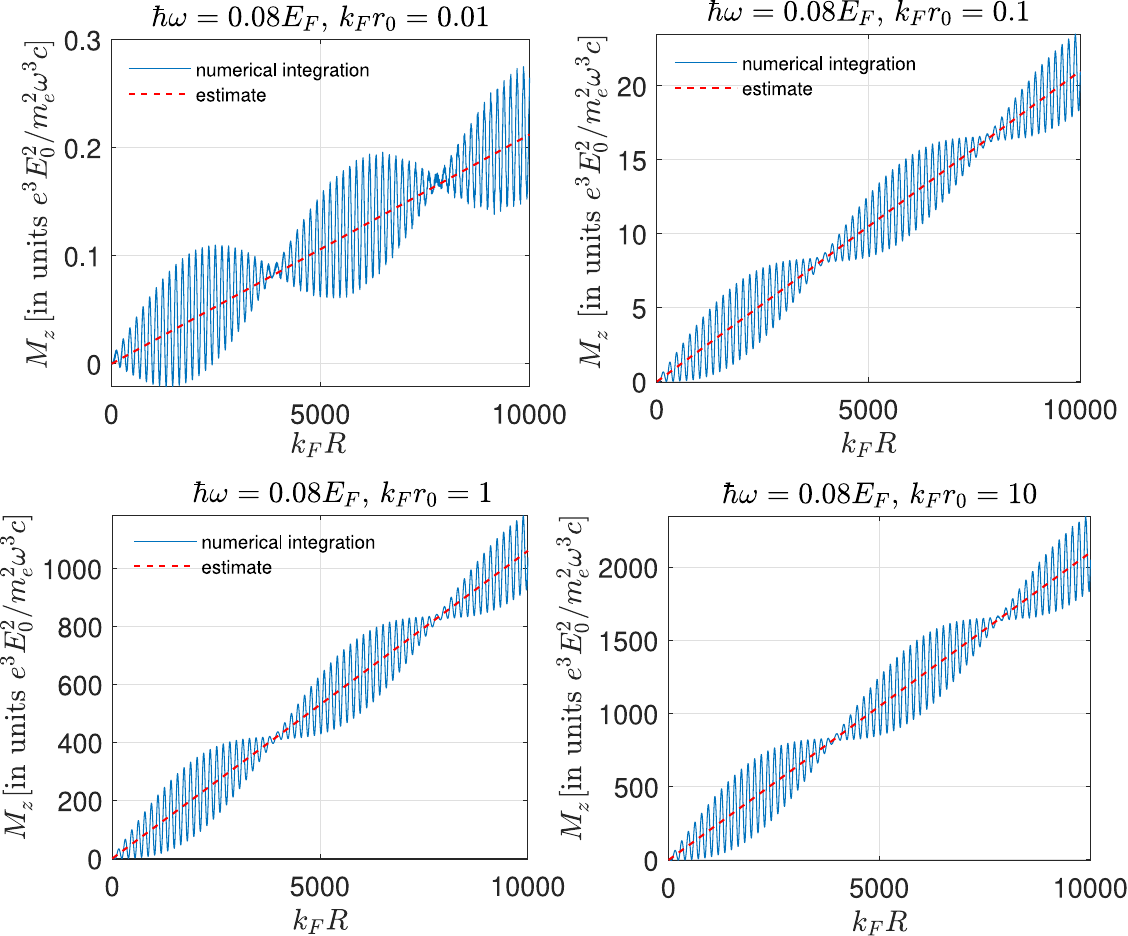}
\caption{Typical plots of the dc magnetic moment $M_z(R)$ defined as magnetization integrated over a sphere of radius $R$ around the impurity. The results were obtained for $\hbar\omega = 0.08E_F$ and for several values of the parameter $k_Fr_0$.}
\label{Fig:3D_illustration}
\end{figure}

Finally, using the above corrections for the wave function, we compute $M_z(R)$. We obtain
\begin{equation}\label{Mz_3D_expression}
M_z(R) = \frac{e}{2m_ec}\left[Q_s(R) + Q_p(R) + I_s(R) + I_p(R)\right] \ ,
\end{equation}
where
\begin{subequations}\label{Mz_3D_contributions}
 \begin{align}
  Q_s = \frac{\hbar e^2E_0^2}{m_e^2\omega^4}\frac{1}{3\pi}\int_0^{k_F}dk\int_0^R dr \ \frac{r^2k^4}{1 + k^2r_0^2}\biggl\{2[j_1(kr)-kr_0y_1(kr)]\left[\frac{k_+^2}{k}r_0y_1(k_+r)-{\rm Im}(q_{0,0,k})\right]\\
  \nonumber
   + \frac{k_+^4}{k^2}r_0^2h_1^{(1)}(k_+r)h_1^{(2)}(k_+r)-|q_{0,0,k}|^2\biggl\} \ ,\\
  Q_p = -\frac{\hbar e^2E_0^2}{m_e^2\omega^4}\frac{2}{3\pi}\int_0^{k_F}dk\int_0^R dr \ k^2r^2j_1(kr){\rm Im}\left\{\frac{kr_0}{1 - ik_+r_0}[k_+^2h_1^{(1)}(k_+r)-k^2h_1^{(1)}(kr)]-q_{1,1,k}^{(2)}\right\} \ ,\\
  I_s = \frac{e^2E_0^2}{m_e\omega^3}\frac{2}{3\pi}\int_0^{k_F} dk\int_0^R dr \ \frac{[j_0(kr)-kr_0y_0(kr)]}{1 + k^2 r_0^2}k^3r^3\left\{2[j_1(kr)-kr_0y_1(kr)]+{\rm Im}\left[q_{0,0,k}(r) + \frac{k_+^2}{k}r_0h_1^{(1)}(k_+r)\right]\right\} \ ,\\
  I_p = \frac{-e^2E_0^2}{m_e\omega^3}\frac{2}{3\pi}\int_0^{k_F}dk \int dr \ k^3r^3 j_1(kr)\left\{2j_0(kr) + {\rm Im}\left[q_{1,1,k}(r) - \frac{k_+r_0}{1-ik_+r_0}h_0^{(1)}(k_+r)\right]\right\} \ .
 \end{align}
\end{subequations}
Typical results of Eqs.~(\ref{Mz_3D_expression}) and (\ref{Mz_3D_contributions}) are shown in Fig.~\ref{Fig:3D_illustration}.

To find analytical expression for the magnetic moment, we neglect oscillating contribution and exponentially decaying ones in the large $R$ limit. As a result, we get the following expression for the magnetic moment
\begin{equation}
 M_z = \frac{\hbar e^3E_0^2}{m_e^3\omega^4 c}\frac{R}{3\pi}\int_0^{k_F}dk\left[\frac{k^2r_0^2(k_+^2 - k_-^2)}{1 + k^2r_0^2} + \frac{k^3k_+r_0^2}{1 + k_+^2r_0^2} - \frac{k^3k_-r_0^2}{1 + k_-^2r_0^2}\right] \ .
\end{equation}
Assuming $\hbar\omega\ll E_F$, we perform the Taylor expansion of the integrands and get 
\begin{equation}
 M_z = \frac{e^3E_0^2}{m_e^2\omega^3c}\frac{2k_FR}{3\pi}\frac{k_F^2r_0^2}{(1 + k_F^2r_0^2)} \ . 
\end{equation}
Corresponding expression for the dc current density has the form
\begin{equation}
\mathbf{j}(\mathbf{r}) = Q_{\rm 3D}\frac{\mathbf{e}_z\times\mathbf{r}}{r^4} \ , \ \ Q_{\rm 3D} = \frac{e^3E_0^2k_F}{2\pi^2m_e^2\omega^3}\frac{k_F^2r_0^2}{1 + k_F^2r_0^2} \ .
\end{equation}
This rectified current generates the magnetic field distribution
\begin{equation}
\mathbf{B} = \frac{4\pi Q_{\rm 3D}}{c}\frac{z\mathbf{r}}{r^4} \ ,
\end{equation}
and the magnetic flux though the plane perpendicular to the $z$-axis
\begin{equation}
\phi = \int d^2\mathbf{r} \ B_z(x,y,z) = \frac{4\pi^2Q_{\rm 3D}}{c} \ .
\end{equation}

\end{document}